\documentclass[
twocolumn,
showpacs,preprintnumbers,amsmath,amssymb]{revtex4}

\begin{document}
\title{Stochastic dynamics beyond the weak coupling limit: thermalization}
\author{A.V. Plyukhin}
\email{aplyukhin@anselm.edu}
 \affiliation{ Department of Mathematics,
Saint Anselm College, Manchester, New Hampshire 03102, USA 
}

\date{\today}

\begin{abstract}
We discuss the structure and asymptotic long-time properties of 
coupled equations for the moments  of a  Brownian
particle's  momentum  $\langle p^n(t)\rangle $ derived microscopically beyond the lowest 
approximation in the weak coupling parameter $\lambda$. 
 Generalized fluctuation-dissipation relations
are derived and shown to  ensure convergence
to thermal equilibrium
to any order in $\lambda$. 

\end{abstract}


\maketitle
\section{Introduction}
Many popular results of non-equilibrium statistical mechanics, such as 
exponential  decay of correlations, hold only on
a time scale much longer than the characteristic relaxation 
time $\tau_b$ for a thermal bath and are thus applicable only for 
sufficiently slow systems with relaxation time $\tau_s\gg\tau_b$.
Such results may be obtained microscopically in the lowest order of  
the perturbation theory with the ratio $\lambda\sim\tau_b/\tau_s$ as a small
parameter and, additionally, 
with a coarse-grained time resolution much larger than
$\tau_b$. 
This approximation is referred to as the weak coupling limit and can be
concisely formulated as a combination of three conditions 
$\lambda\to 0$, $t\to\infty$, with $\lambda t$ finite.

There are, of course,  many situations of physical interest 
when the weak coupling
approximation does not suffice~\cite{Broeck,Kosov,Froese}. 
Van Kampen developed a highly successful theory which allows us to 
take into account terms of higher orders in $\lambda$, but  still 
using as a prerequisite  the time coarse graining in a form of the assumption
that the system interacts with the bath via 'instantaneous' 
binary collisions~\cite{Kampen}. 
The relaxing of this rather  artificial limit leads, in general, to  
non-Markovian master or Langevin equations which 
are more difficult to handle than their Markovian counterparts.
Beyond the weak coupling limit, these  equations have a rather complicated
structure~\cite{Kapral, Plyukhin2}, 
and not much is known about their properties.
In particular, while van Kampen's theory is consistent 
with equilibrium statistical mechanics~\cite{Plyukhin1}, 
the relaxation  to Maxwell-Boltzmann equilibrium
within a more general
approach, which takes into account multiple collisions and non-Markovian 
effects,
is not entirely obvious and was 
questioned in several 
studies~(see~\cite{Grigolini,Plyukhin2} and references therein). 


One purpose of this paper is to put the equations of stochastic
dynamics into a form convenient for the evaluation of stationary solutions
to any order in the weak coupling parameter $\lambda$. 
Most previous works in this direction seek to generalize the  Fokker-Planck 
equation for the distribution function $f(a,t)$ for a targeted dynamical 
variable $a$. One difficulty with this
approach is that  beyond the lowest order in $\lambda$ generalized
Fokker-Planck equations involve derivatives $\partial^n f/\partial a^n$ 
of orders higher than two and do not guarantee positive definiteness 
of the solution. Also, and perhaps more importantly, within this approach 
it is not clear how to, in a systematic perturbative way, 
take into account non-Markovian effects.
The same perturbation technique
which justifies  results in the weak coupling limit may not
work well in higher orders in $\lambda$, leading to 
stationary distributions for the system  with  
an equilibrium temperature different than that of the bath~\cite{Plyukhin2}. 
Neither real~\cite{Li} nor numerical~\cite{Shin}
experiments suggest such a possibility. Some other troubles with 
non-Markovian Fokker-Planck equations were reported in~\cite{Mazo}.
In this paper we show that 
some of these difficulties can be avoided if one  works with microscopically
derived  Langevin equations for
the powers $a^n$ of a targeted variable. These equations can be readily used
to derive equations for the moments $\langle a^n(t)\rangle$ which are linear
and not
difficult to work with even in non-Markovian form, at least as far as 
stationary solutions are concerned.

Although the discussion can be carried on a very general level, we choose 
to consider,
for the sake of better visualization, the  archetype example 
of a Brownian particle of 
mass $M$ immersed in a infinitely large bath at temperature $T=1/\beta$ and
composed of molecules with mass $m\ll M$.  
We will employ the method by Albers et al.~\cite{Albers} 
to derive  
equations of motion for the moments  $\langle p^n(t)\rangle$ of the 
Brownian particle's (scaled)  momentum. It  is shown that to any order 
in the weak coupling parameter $\lambda=(m/M)^{1/2}$ the moments relax to
the equilibrium values prescribed  by equilibrium statistical mechanics. 
Convergence to thermal equilibrium is  guided by generalized
fluctuation-dissipation relations, which can hardly be derived by
any other method but microscopically.  
On the other hand, thermalization is found to be insensitive 
to particular relations between involved microscopic correlation functions. 
This leads to the  optimistic
conclusion that a consistent theory need not be totally microscopic.

\section{Exact equations}
Consider a system of  $N$ bath molecules of mass $m$  interacting with each
other  and with a  Brownian particle of mass $M$ via short range potential 
$U$.
The Hamiltonian of  the system is 
$H=P^2/2M+H_0(x)$,
where $P$ is the momentum of the particle, and 
$H_0(x)$ 
is the Hamiltonian of bath molecules in the field of the particle fixed 
at the position $x$.
Introducing as usual the scaled momentum of the particle
$p=\lambda P$ with $ \lambda=(m/M)^{1/2}$,
the Liouville operator of the systems can be written as  
\begin{eqnarray}
L=L_0+\lambda L_1
\end{eqnarray}
where the operator 
$L_1$ acts on the particle's variables only
\begin{eqnarray}
L_1=\frac{p}{m}\frac{\partial }{\partial x}+F\frac{\partial }{\partial p},
\end{eqnarray}
while 
$L_0$ governs the dynamics of the bath in the field of the 
fixed particle,
\begin{eqnarray}
L_0=\sum_{i=1}^N \left\{ \frac{p_i}{m}\frac{\partial }{\partial x_i}
+F_i\frac{\partial }{\partial p_i} \right\},
\end{eqnarray}
and thus satisfies the relation $L_0H_0=0$.
In these equations  $x_i$, $p_i$ are coordinates and momenta  of 
bath molecules, 
$F_i=-\partial U/\partial x_i$  and $F=-\partial U/\partial x$ 
are the forces on the {\it i}th molecules and on the particle, respectively.

We wish to decompose the exact equation 
\begin{eqnarray}
\frac{d}{dt} p^n(t)=e^{Lt} L \, p^n,
\label{start}
\end{eqnarray}
where $p^n=p^n(0)$,  into a form convenient to derive the Langevin 
equation for $p^n$ 
using an expansion in the small parameter $\lambda$. 
Using the operator identity
\begin{eqnarray}
e^{\mathcal A t }=
e^{(\mathcal{A+B})t}-\int_0^t  d\tau e^{\mathcal A(t-\tau)}\mathcal B e^{(\mathcal{A+B})\tau},
\label{identity_0}
\end{eqnarray}
with $\mathcal A=L$, $\mathcal B=-\mathcal P L$, and $\mathcal Q=
1-\mathcal P$ one gets
\begin{eqnarray}
e^{Lt}=
e^{\mathcal Q Lt}+\int_0^t  d\tau e^{L(t-\tau)}\mathcal P L e^{\mathcal Q L\tau}.
\label{identity}
\end{eqnarray}
The operator  $\mathcal P$ is convenient to chose to be 
a projector operator ($\mathcal P^2=\mathcal P$) that averages  
over initial values of bath variables
\begin{eqnarray}
\mathcal P A=\langle A\rangle =\int \rho \,A\, dx_1... dx_N dp_1 ... dp_N
\end{eqnarray}
with the canonical distribution
\begin{eqnarray}
\rho=\frac{1}{Z} \,e^{-\beta H_0}.
\end{eqnarray}
Such defined projection operator $\mathcal P$ and its complement
$\mathcal Q=1-\mathcal P$ 
satisfy the relations
\begin{eqnarray}
\mathcal P L_0=0,\qquad \mathcal Q L_0=L_0.
\label{ort}
\end{eqnarray}
With  (\ref{identity}) and (\ref{ort}), the right-hand side of (\ref{start})
can be written as  
\begin{eqnarray}
e^{Lt}\mathcal (Lp^n)=e^{\mathcal Q Lt}\mathcal ( Lp^n)+
\lambda \int_0^t  d\tau e^{L(t-\tau)}\mathcal P L_1 e^{\mathcal Q L\tau} (L p^n).
\nonumber
\end{eqnarray}
Then (\ref{start}) takes the desirable pre-Langevin form
\begin{eqnarray}
\frac{d}{dt}p^n(t)=
\lambda K_n(t)+\lambda^2 \int_0^t d\tau \,e^{L(t-\tau)}\mathcal P L_1
K_n(\tau), 
\label{exact}
\end{eqnarray}
where
\begin{eqnarray}
K_n(t)=\lambda^{-1} e^{\mathcal QLt}L p^n
\label{force}
\end{eqnarray}
plays the role of a "random" (rapidly fluctuating) force.

Notice that the above expression for $K_n(t)$ can be 
alternatively written with an
additional factor $\mathcal Q=1-\mathcal P$: 
\begin{eqnarray}
K_n(t)=\lambda^{-1} e^{\mathcal Q Lt}\mathcal Q L p^n.
\label{force2}
\end{eqnarray}
This is because  
\begin{eqnarray}
\mathcal P L p^n=\lambda \mathcal P L_1 p^n=
\lambda n \,p^{n-1}\mathcal P F=\lambda n \, p^{n-1} \langle F\rangle=0.
\nonumber
\end{eqnarray}
The form (\ref{force2}) makes it obvious that the fluctuating term $K(t)$ 
is zero centered
\begin{eqnarray}
\langle K_n(t)\rangle=\mathcal P K_n(t)\sim \mathcal {P\,Q}=0.
\end{eqnarray}
Another useful identity involving $K_n(t)$,  which can be readily proved 
by integrating by parts, reads
\begin{eqnarray}
\left\langle \frac{\partial}{\partial x} K_n(t)\right\rangle=
\mathcal P  \frac{\partial}{\partial x} K_n(t)=-\beta \langle FK_n(t)\rangle.
\label{aux}
\end{eqnarray}

Equipped with these relations, one  eventually write Eq. (\ref{exact}) 
in the form
\begin{eqnarray}
\frac{d}{dt}p^n(t)&=&\lambda K_n(t)
\label{exact2}\\
&+&\lambda^2 \int_0^t
d\tau \,e^{L(t-\tau)}
\left( \frac{\partial}{\partial p}-\frac{\beta\,p}{m}
\right)
\langle  F\,K_n(\tau)\rangle 
\nonumber
\end{eqnarray}
with the fluctuating force
\begin{eqnarray}
K_n(t)=n \, e^{\mathcal Q L t}\, F\,p^{n-1}.
\label{K}
\end{eqnarray}

The procedure outlined above is generic and can be easily generalized to
derive exact Langevin-like equation of motion for 
an arbitrary dynamical variable or distribution function~\cite{Albers}.

\section{Lowest-order perturbation}
As it is, the exact equation (\ref{exact2}) is  of little help because it
contains the variables of interest $p^n$  implicitly in the operator 
$e^{\mathcal Q L t}$.  In order to make this dependence explicit one can expand 
$e^{\mathcal Q L t}=e^{L_0t+\lambda\mathcal QL_1t}$
in powers of $\lambda$ iteratively using the relation (\ref{identity_0})
with $\mathcal A=L_0$ and $\mathcal B=\lambda \mathcal Q L_1$:
\begin{eqnarray}
\!\!\!\!\!\!&&e^{\mathcal QLt}=e^{L_0 t}+
\lambda\int_0^t\, d\tau e^{L_0(t-\tau)}
\mathcal Q L_1 e^{L_0 \tau}
\label{expansion}\\
\!\!\!\!\!\!&&+\lambda^2
\int_0^t \!\!\!\!d\tau_1 \!\!\int_0^{\tau_1} \!\!\!d\tau_2\,
e^{L_0(t-\tau_1)}\mathcal Q L_1 
e^{L_0(\tau_1-\tau_2)}\mathcal Q L_1 e^{L_0\tau_2} + ... 
\nonumber 
\end{eqnarray}
Substitution of this into Eq.(\ref{K}) for the random force $K_n(t)$  
leads to the expansion
\begin{eqnarray}
K_n(t)=\sum_{i=0}^\infty \lambda^i \, K_n^{(i)} (t).
\label{expansion2}
\end{eqnarray}
The lowest order term reads
\begin{eqnarray}
K_n^{(0)}(t)=n p^{n-1} F_0(t),
\end{eqnarray} 
where $F_0(t)=e^{L_0t}F$ is the force exerted by the bath on the fixed particle, which can be called the pressure force. 
As follows from (\ref{K}) and (\ref{expansion}), the higher order terms $K_i$
can be obtained  recurrently as follows
\begin{eqnarray}
K_n^{(i)}(t)=\int_0^t d\tau e^{L_0 (t-\tau)} \mathcal Q L_1\, K_n^{(i-1)}(\tau).
\label{recurrent}
\end{eqnarray}
Note that each term in the expansion of the random force 
is zero-centered, $\langle K_n^{(i)}(t)\rangle=0$.

In the lowest order in $\lambda$ the exact equation of motion  (\ref{exact2})   
takes the form of the generalized Langevin equation
\begin{eqnarray}
&&\frac{d}{dt} p^n(t)=\lambda n p^{n-1}F_0(t)\label{gle0}\\
&&-\lambda^2\,\frac{\beta}{m}\int_0^t d\tau \,c_0(t-\tau)\, p^n(\tau)
\nonumber\\
&&+\lambda^2 (n-1) \int_0^t d\tau \, c_0(t-\tau)\, p^{n-2}(\tau), 
\nonumber
\end{eqnarray} 
where the memory kernel $c_0(t)$ is given by the correlation function of the
pressure force,
\begin{eqnarray}
c_0(t)=n\langle FF_0(t)\rangle.
\end{eqnarray} 
Taking the average of Eq. (\ref{gle0}) one obtains 
for the moments
\begin{eqnarray}
A_n(t)=\langle p^n(t)\rangle=\mathcal P p^n(t)
\end{eqnarray} 
the following equation
\begin{eqnarray}
&&\frac{d}{dt} A_n(t)=
-\lambda^2\,\frac{\beta}{m}\int_0^t d\tau \,c_0(t-\tau) A_n(\tau)
\label{nm0} \\
&&+\lambda^2 (n-1) \int_0^t d\tau \, c_0(t-\tau) A_{n-2}(\tau). 
\nonumber
\end{eqnarray} 
In Markovian limit it takes the familiar form~\cite{Coffey} 
\begin{eqnarray}
\!\!\!\!\!\!
\frac{d}{dt} A_n(t)=
-\lambda^2\,n\,\gamma_0 A_n(t)+
\lambda^2 n (n-1) \frac{m}{\beta} \gamma_0 A_{n-2}(t)
\label{m0}
\end{eqnarray} 
with the damping coefficient 
\begin{eqnarray}
\gamma_0=\frac{\beta}{n\,m}\int_0^\infty \!dt\,  c_0(t)=
\frac{\beta}{m}\int_0^\infty \!dt\,  \langle F F_0(t)\rangle.
\end{eqnarray}
Of course, Eq.  (\ref{m0}) can be derived more easily
from the phenomenological Langevin equation 
\begin{eqnarray}
\dot p(t)=-\lambda^2\gamma_0 \, p(t)+\lambda F_0(t)
\end{eqnarray}
under the assumption that $F_0(t)$ is Gaussian noise~\cite{Coffey}.
An important outcome  of the above microscopic 
derivation is that it shows that the assumption of Gaussian random force  
is in fact unnecessary. Another advantage of the non-Markovian equation 
(\ref{nm0}) is that it holds for any time, while its Markovian counterpart 
(\ref{m0}) applies only  on a time  scale longer than the characteristic 
time for the decay of the correlation function $c_0(t)$.

It is  easy to show that  both Markovian (\ref{m0}) 
and non-Markovian  (\ref{nm0}) equations 
describe relaxation  of the moments  to equilibrium values
$A_n^e$ prescribed by equilibrium statistical mechanics, in particular
\begin{eqnarray}
A_2^{e}=\frac{m}{\beta}, \quad A_4^{e}=3 \left(\frac{m}{\beta}\right)^2, 
\quad 
A_6^{e}=15 \left(\frac{m}{\beta}\right)^3,
\label{eqvalues}
\end{eqnarray}
and in general for even $n$
\begin{eqnarray}
A_n^{e}=(n-1)\,\frac{m}{\beta}\,A_{n-2}^{e}
\label{eqvalues_gen}
\end{eqnarray}
(equilibrium odd moments $A_{2n+1}^e$ vanish due to symmetry). 
For non-Markovian equation 
(\ref{nm0}), thermalization can be proved using Laplace transformation
$\tilde A_n(s)=\int_0^\infty dt e^{-st}A_n(t)$,
\begin{eqnarray}
\tilde A_n(s)&=&\frac{A_n(0)}{s+\lambda^2(\beta/m)\,\tilde c_0(s)}
\label{laplace}
\\
&+&
\frac{\lambda^2(n-1)\,\,\tilde c_0(s)}{s+\lambda^2(\beta/m)\,\tilde c_0(s)}
\,\,\tilde  A_{n-2}(s),
\nonumber
\end{eqnarray}
where $\tilde A_0(s)=1/s$. For asymptotic long-time values
\begin{eqnarray}
A_n^e=\lim_{t\to \infty} A_n(t)=\lim_{s\to 0} s \tilde A_n(s)
\end{eqnarray}
Eq.(\ref{laplace})  gives the thermal equilibrium result (\ref{eqvalues_gen}).

It also can be seen from from (\ref{laplace}) that relaxation  
to thermal equilibrium does  not occur if 
\begin{eqnarray}
\tilde c_0(s)\sim s^\delta, \quad \delta\ge 1, \quad  \mbox{as} 
\quad s\to 0. 
\label{ne_condition}
\end{eqnarray}
In this case the damping coefficient $\gamma_0$ vanishes 
\begin{eqnarray}
\gamma_0=\frac{\beta}{n\,m}\int_0^\infty dt \, c_0(t)=
\frac{\beta}{n\,m}\,\tilde c_0(0)=0,
\end{eqnarray}
which corresponds to superdiffusion of the particle~\cite{superdiffusion}. 
Relation
(\ref{ne_condition}) as a condition of 
non-ergodic relaxation of the second moment $A_2(t)$ was discussed
in~\cite{Morgado}. As we see, in the lowest order 
in $\lambda$ the same condition holds for higher moments as well.

\section{Higher-order results} 
As follows from (\ref{exact2}), higher-order terms 
in the $\lambda$-expansion 
of the fluctuating force
\begin{eqnarray}
K_n(t)= K_n^{(0)}(t)+\lambda\,K_n^{(1)}(t)+\lambda^2\,K_n^{(2)}(t)+\ldots
\end{eqnarray}
appear in the
equation for the moments $A_n$ in the form of correlations 
$\langle F K_n^{(i)}(t)\rangle$. Evaluation of these 
correlations may be  discouragingly
complicated even for 
simplified models~\cite{Plyukhin3}. 
However, as we show  in this section, 
a detailed evaluation of 
microscopic correlations is unnecessary
to  demonstrate convergence to thermal equilibrium 
to any perturbation order. All one actually needs is 
to find  the explicit dependence of correlations 
$\langle F K_n^{(i)}(t)\rangle$ 
on the particle's momentum $p$, 
which is a much easier task.

First, recall that in the lowest order $K_0(t)=np^{n-1}F_0(t)$, so 
\begin{eqnarray}
\langle F K_n^{(0)}(t)\rangle=c_0(t)\,p^{n-1},
\label{kernel0}
\end{eqnarray}
with $c_0(t)=n\langle FF_0(t)\rangle$. Substitution of this into the exact
equation (\ref{exact2}) and taking the average leads
to an equation for the moments in the form (\ref{nm0}).

Next, it follows from the recurrence relation (\ref{recurrent}) that
$K_n^{(i+1)}\sim L_1K_n^{(i)}=
[(m^{-1}\partial/\partial x)\,\,p+F\,\,\partial/\partial p]\,K_n^{(i)}$. 
Therefore
the correlations
$\langle FK_n^{(i)}(t)\rangle$ as functions of $p$  can be obtained
recurrently  as follows 
\begin{eqnarray}
\langle F K_n^{(i+1)}(t)\rangle
\sim
\left(p+\partial/\partial p\right)
\langle F K_n^{(i)}(t)\rangle.
\label{aux2}
\end{eqnarray}
From (\ref{kernel0}) and (\ref{aux2}) one obtains
\begin{eqnarray}
\!\!\!\!\!\langle F K_n^{(1)}(t)\rangle =c_{10}(t)\,p^n+
c_{11}(t)\, p^{n-2}.
\label{kernel1}
\end{eqnarray}
Explicit evaluation of the functions $c_{10}(t)$ and $c_{11}(t)$
(see Appendix) immediately reveals that for a homogeneous bath both functions 
vanish identically. More generally, it can be proved with the
standard symmetry argument~\cite{Berne} that 
\begin{eqnarray}
\langle F K_n^{(i)}(t)\rangle=0  \quad \mbox{for odd} \quad i.
\label{kernel_odd}
\end{eqnarray} 
Then the first non-vanishing correction 
to the kernel $\langle FK_n^{(0)}(t)\rangle$ 
is $\lambda^2\langle FK_n^{(2)}(t)\rangle$. 
From (\ref{kernel1}) and (\ref{aux2}) one gets
\begin{eqnarray}
\!\!\!\!\!\langle F K_n^{(2)}(t)\rangle =c_{20}(t)\,p^{n+1}+
c_{21}(t)\, p^{n-1}+c_{22}(t)\, p^{n-3}.
\label{kernel2}
\end{eqnarray}
Explicit expressions 
for the functions  $c_{2i}(t)$ 
are not needed for our purposes, yet for the sake of completeness 
they are given in the Appendix. It is helpful, however, to notice that
the function $c_{22}(t)$ involves the factor $(n-1)(n-2)$, which makes it
vanish for $n<3$, so that the above expression 
involves only positive powers of $p$.
The same is true for higher order corrections. The next non-zero
term has the form  
\begin{eqnarray}
\langle F K_n^{(4)}(t)\rangle &=&
c_{40}(t)\,p^{n+3}+c_{41}(t)\, p^{n+1}
\label{kernel4}\\
&+&c_{42}(t)\, p^{n-1}+c_{43}(t)\,p^{n-3}+c_{44}(t)\,p^{n-5},
\nonumber
\end{eqnarray}
which can be obtained by applying twice the recurrence 
relation (\ref{aux2}) to the correlation $\langle F K_n^{(2)}(t)\rangle$ given
by (\ref{kernel2}).  The functions $c_{ij}(t)$ also depend on $n$.
One can show that the above expression 
involves only positive powers of $p$, since
$c_{43}\sim\prod_{k=1}^3(n-k)$ and $c_{44}\sim \prod_{k=1}^5(n-k)$.

With the pattern given by the above relations, the 
general expression can be written in the form
\begin{eqnarray}
\langle F K_n^{(i)}(t)\rangle =p^{n-1}\sum_{j=0}^i c_{ij}(t)\, p^{i-2j},
\end{eqnarray}
where $c_{ij}(t)$ vanish identically for odd $i$ due
to symmetry, and $c_{00}(t)\equiv c_0(t)$. 
We see that the higher the order of perturbation $i$, the larger the number of 
powers $p^i$ to which the variable of interest $p^n$ is coupled to.  

Equipped with the above  relations, one can write the equation 
for the moments $A_n(t)=\langle p^n(t)\rangle$ to any
order in $\lambda$. As an example, let us consider the equations for the
second moment $A_2$. In order to get the first non-zero correction 
to the lowest-order results discussed in the
previous section, we need to expand the fluctuating force $K_2(t)$ up to 
order $\lambda^2$,
\begin{eqnarray}
K_2(t)= K_2^{(0)}(t)+\lambda\,K_2^{(1)}(t)+\lambda^2\,K_2^{(2)}(t).
\end{eqnarray}
Then from (\ref{kernel0}), (\ref{kernel_odd}) and (\ref{kernel2}) 
with we get
\begin{eqnarray}
\langle F K_2(t)\rangle = c_0(t)\,p+\lambda^2 \left\{
c_{20}(t)\,p^3+c_{21}(t)\,p\right\}.
\label{aux55}
\end{eqnarray}
Substituting this into the exact equation (\ref{exact2})  one obtains 
the following equation
\begin{eqnarray}
&&\!\!\!\!\!\!\!\frac{d}{dt} A_2(t)=\lambda^2\left[
1-(\beta/m)\,A_2\right]\circ c_0
\label{A2}\\
&&\!\!\!\!\!\!\! +\lambda^4\left\{
\left[3\,c_{20}-(\beta/m) \,c_{21}\right]\circ A_2
-(\beta/m)\,c_{20}\circ A_4+\,c_{21}\circ 1
\right\}. 
\nonumber 
\end{eqnarray}
Here and below  we adopt 
the shorthand notation $f\circ g$ for the convolution integral 
$\int_0^t d\tau f(\tau)g(t-\tau)$.
One observes that to the given order $A_2$ is coupled to $A_4$, which is 
in contrast with the lowest-order approximation where the equation for
$A_2$ is closed.

Applying  Laplace transformation one can write  
the long-time stationary value for $A_2$ as a fraction
\begin{eqnarray}
\lim_{t\to\infty} A_2(t)=\lim_{s\to 0}s\tilde A_2(s)=\frac{N}{D}
\label{limit4}
\end{eqnarray}
with  
the denominator
\begin{eqnarray}
D=\lambda^2(\beta/m)\,\tilde c_0(0)+\lambda^4 (\beta/m) \,\tilde c_{21}(0)
-3\lambda^4 \, \tilde c_{20}(0)
\nonumber
\end{eqnarray}
and the numerator
\begin{eqnarray}
N=\lambda^2 \tilde c_0(0)+\lambda^4 \tilde c_{21}(0)-
\lambda^4 \,(\beta/m) \,\tilde c_{20}(0) \,\lim_{s\to 0} s\tilde A_4(s).
\nonumber
\end{eqnarray}
The stationary value for the fourth moment 
$\lim_{s\to 0}s\tilde A_4(s)=\lim_{t\to\infty} A_4(t)$
appears here multiplied by $\lambda^4$, 
and therefore the value $3(m/\beta)^2$
found in the lowest-order limit, Eq. (\ref{eqvalues}), should be assigned 
to it. Then (\ref{limit4}) gives for the second moment 
the same equilibrium value as in the lowest perturbation order, 
$A_2(t)\to A_2^e=m/\beta$.

Higher moments can be handled in a similar way. For instance, as follows from
(\ref{kernel0}), (\ref{kernel_odd}) and (\ref{kernel2}),
for the fourth
moment $A_4=\langle p^4\rangle$
the kernel $\langle F K_4(t)\rangle$ to order $\lambda^2$ takes  the form 
\begin{eqnarray}
\langle F K_4(t)\rangle = c_0\,p^3+\lambda^2 \left\{
c_{20}\,p^5+c_{21}\,p^3+c_{22}\,p\right\}.
\end{eqnarray}
Substitution of this into (\ref{exact2}) gives the equation
\begin{eqnarray}
&&\!\!\!\!\!\!\!\frac{d}{dt} A_4(t)=\lambda^2\left[
3c_0\circ A_2-(\beta/m)\,c_0\circ A_4\right]
\label{A4}\\
&&\!\!\!\!\!\!\! +\lambda^4
\Bigl\{
\left[3\,c_{21}-(\beta/m) \,c_{22}\right]\circ A_2
+\left[5\,c_{20}-(\beta/m) \,c_{21}\right]\circ A_4\nonumber\\
&&\quad-(\beta/m)\, c_{20}\circ A_6
+c_{22}\circ 1
\Bigr\}.
\nonumber 
\end{eqnarray}
Applying Laplace transform and recalling that in the lowest order
$\lim_{s\to 0} s\tilde A_6(s)=A_6^e=15\,(m/\beta)^3$, one finds
\begin{eqnarray}
\lim_{t\to\infty} A_4(t)=\lim_{s\to 0} s \tilde A_4(s)=A_4^e=3\,(m/\beta)^2,
\end{eqnarray}
which is again the equilibrium result which we 
already obtained in the lowest order.

No new features appear as one extends the technique to higher perturbation 
orders. The next non-zero correction  corresponds to the
expansion of the fluctuating  force to order $\lambda^4$, 
$K_n(t)=\sum_{i=0}^4\lambda^i\,K_n^{(i)}(t)$. The correlations
$\langle F K_n^{(i)}(t)\rangle$ for $i=0,2,4$ are given by equations
(\ref{kernel0}), (\ref{kernel2}), and (\ref{kernel4}), respectively.
For example,  for the second
moment $A_2$ one obtains
\begin{eqnarray}
\langle F K_2(t)\rangle &=&c_0\,p+\lambda^2\left\{
c_{20}\, p^3+c_{21}\,p\right\}\\
&+&\lambda^4\left\{
c_{40}\,p^5+c_{41}\,p^3+c_{42}\,p\right\}.\nonumber
\end{eqnarray}
Then substitution into (\ref{exact2}) leads to
an equation which differs from Eq.(\ref{A2}) by the presence of terms of
order $\lambda^6$,  
\begin{eqnarray}
&&\!\!\!\!\!\!\frac{d}{dt} A_2(t)=\lambda^2\left[
1-(\beta/m)\,A_2\right]\circ c_0\\
&&\!\!\!\!\!\! +\lambda^4\Bigl\{
[3\,c_{20}-(\beta/m) \,c_{21}]\circ A_2
-(\beta/m)\,c_{20}\circ A_4+\,c_{21}\circ 1
\Bigr\} 
\nonumber
\\
&&\!\!\!\!\!\!+\lambda^6\Bigl\{
[3\,c_{41}-(\beta/m) \,c_{42}]\circ A_2
+[5\,c_{40}-(\beta/m) \,c_{41}]\circ A_4\nonumber\\
&&\quad\,\,\,-(\beta/m)\,c_{40}\circ A_6+\,c_{42}\circ 1
\Bigr\}. 
\nonumber 
\end{eqnarray}
Applying Laplace transformation and
assigning equilibrium values found to lower perturbation orders 
for long-time limits of $A_4$ and $A_6$,
one again obtains $A_2(t)\to A_2^e=m/\beta$.

\section{Conclusion} 

In the weak coupling
limit, the equations for the first two moments $A_1=\langle p\rangle $ and 
$A_2=\langle p^2\rangle$ of the Brownian particle's momentum  
are closed, while  higher 
moments $A_n$ are coupled to $A_{n-2}$ only, see Eq. 
(\ref{m0}). To higher orders in the weak-coupling parameter $\lambda$,
a larger number of moments are coupled.
The higher perturbation order, the larger the 
number of different moments appear in the  equation for $A_n$.  
In a Markovian limit the  equations coincide with those obtained by van Kampen
within the  instantaneous binary collision model,
but contain parameters expressed in a 
totally microscopic way.
For example, according to (\ref{A2}),  to order $\lambda^4$
the equation for $A_2$  is not closed but involves coupling to $A_4$:
\begin{eqnarray}
\frac{d}{d t}A_2(t)=-\gamma_1 A_2(t)-\gamma_2 A_4 +\gamma_3.
\label{A2_mark}
\end{eqnarray}
Dissipative coefficients $\gamma_i$ are given by 
fluctuation-dissipation relations
\begin{eqnarray}
\gamma_1&=&\lambda^2(\beta/m)\, \alpha_0
-3\lambda^4\,\alpha_{20}+\lambda^4(\beta/m)\,\alpha_{21},
\nonumber
\\
\gamma_2&=&\lambda^4(\beta/m)\,\alpha_{20},\label{FDT}\\
\gamma_3&=&\lambda^2 \,\alpha_0+\lambda^4 \,\alpha_{21},\nonumber
\end{eqnarray}
Here, coefficients 
$\alpha_0=\int_0^\infty c_0(t) \,dt$ and 
$\alpha_{2i}=\int_0^\infty c_{2i}(t)\,dt$
are system-dependent parameters, 
given by integrals of microscopic correlations.

Since more than one microscopic parameters $\alpha$  are involved in 
(\ref{FDT}), 
it appears natural to ask whether any  constraints on
their relations do exist
which  ensure relaxation of the system to thermal
equilibrium with the bath. 
The present paper shows that
the system's thermalization is guaranteed by 
fluctuation-dissipation relations alone
and no additional 
relations between microscopic parameters are required. 
Convergence to thermal equilibrium with the bath occurs 
to any order in $\lambda$, in both Markovian and non-Markovian regimes.  
For instance, 
given the asymptotic result $A_4(t)\to A_4^e=3 (m/\beta)^2$ 
found in the weak coupling limit,  Eqs.(\ref{A2_mark}) and 
(\ref{FDT}) give for $A_2$ in the long-time limit
the equilibrium value 
$A_2(t)\to(\gamma_3-\gamma_2 \,A_4^{e})/\gamma_1=m/\beta$.

That 
thermalization puts no constraints on microscopic parameters $\alpha$  opens an
attractive avenue for phenomenological modeling. One cannot use, say,  
Eq. (\ref{A2_mark}) with arbitrary postulated values for coefficients 
$\gamma_i$ since
such an equation in general would disagree with equilibrium statistics.  
On the other hand, Eq.(\ref{A2_mark}) supplemented with fluctuation-dissipation 
relations (\ref{FDT}) for $\gamma_i$ with arbitrary $\alpha$ 
is thermodynamically consistent.

Although our attention here was
focused on the issue of thermalization and consistency with equilibrium
statistical mechanics, most interesting applications of 
the developed formalism are expected, 
of course, for time dependent phenomena. 
The coupling of a larger number of moments may result in much richer
dynamics compared to that in the weak coupling limit.
For a Markovian limit 
this was illustrated in {\cite{Broeck,Kosov,Froese}, but results from these 
studies 
may be obtained (and in fact, most of them were) within the framework of 
more simple
van Kampen theory.    For future studies, it would be 
interesting to identify situations where the
non-Markovian form of the equations obtained in this paper would be 
essential and responsible for qualitatively new features.  
Application to Kramers' activated escape problem seems particularly promising,
considering the recent demonstration 
that non-linear corrections to the dissipative force, which are 
of higher orders in $\lambda$,  
may be important 
in the underdamped regime~\cite{Kosov}.

There are several limitations of the presented study.
One  inevitable loophole is the tacit  assumption that each
term in the $\lambda$-expansion is bounded for all time. 
To the best of our knowledge a general proof of this is still lacking. 
Also, we have assumed that the relative smallness of terms is determined
solely by their dependence on $\lambda$. For instance, in (\ref{FDT})  
$\gamma_2\sim\lambda^4$ is assumed to be smaller 
than $\gamma_1\sim\lambda^2$. This is not necessarily true since 
$\gamma_1$ involves factors which
are integrated
correlation functions and may vanish identically or be very small.
In such situations the system may exhibit non-ergodic behavior (does not 
thermilize to the bath temperature) as discussed  elsewhere~\cite{Morgado}.
A more exotic condition of ergodicity breaking is considered 
in~\cite{Plyukhin4}.

\onecolumngrid
\begin{acknowledgments}

This work was supported by a Saint Anselm College summer research grant.
Encouraging discussions with Gregory Buck and Stephen Shea are appreciated.
\end{acknowledgments}

\appendix*
\section{}
Exact explicit expressions for the functions $c_{ij}(t)$ 
in Eqs. (\ref{kernel1}) and 
(\ref{kernel2}) can be written respectively as
\begin{eqnarray}
c_{1j}(t)=\int_0^td\tau\,C_{1j}(t, \tau),\qquad
c_{2j}(t)=\int_0^td\tau_1\int_0^{\tau_1}d\tau_2 \,C_{2j}(t, \tau_1, \tau_2)
\nonumber
\end{eqnarray}
where correlation functions $C_{ij}$ read
\begin{gather*}
C_{10}=\frac{1}{m}\,\left\{
\left
\langle F e^{L_0(t-\tau)}\frac{\partial F_0(\tau)}{\partial x}\right\rangle
-\langle F\rangle \left\langle
\frac{\partial F_0(\tau)}{\partial x}\right\rangle\right\},
\nonumber \\
C_{11}=(n-1)\left\{
\langle F F_0(t-\tau)F_0(t)\rangle
-\langle F\rangle\langle FF_0(\tau)\rangle
\right\},\nonumber \\
C_{20}=\frac{1}{m^2}
\left\langle
F e^{L_0(t-\tau_1)}\frac{\partial}{\partial x}  
e^{L_0(\tau_1-\tau_2)} \frac{\partial}{\partial x} F_0(\tau_2)
\right\rangle, \nonumber\\
C_{21}=\frac{n-1}{m}
\left\langle
F e^{L_0(t-\tau_1)}\frac{\partial}{\partial x} [F_0(\tau_1-\tau_2)F_0(\tau_1)]
\right\rangle
+
\frac{n}{m}\left\langle
F  F_0(t-\tau_1)e^{L_0(t-\tau_2)}\frac{\partial}{\partial x}F_0(\tau_2)
\right\rangle
+\frac{\beta}{m}\langle F F_0(t-\tau_1)\rangle 
\left\langle F F_0(\tau_2)
\right\rangle,\nonumber\\
C_{22}=(n-1)(n-2)\Bigl\{
\langle F F_0(t-\tau_1) F_0(t-\tau_2) F_0(t)\rangle
-\langle F F_0(t-\tau_1)\rangle \langle  F F_0(\tau_2)\rangle
\Bigr\}.
\nonumber
\end{gather*} 
Here $F=F(0)=F_0(0)$. For  a homogeneous bath $C_{10}=C_{11}=0$ by symmetry. 
The identity (\ref{aux}) was used for the derivation of
the last term of $C_{21}$.
Explicit evaluation of these and similar 
functions for a particular model is discussed
in \cite{Plyukhin3}.

\twocolumngrid


\end{document}